\documentclass[12pt]{article}

\usepackage{sbc-template}

\usepackage{graphicx,url}

\usepackage[brazil]{babel}   
\usepackage[utf8]{inputenc}  
\usepackage{xspace}
\usepackage{subfigure}
\usepackage{multirow}

\title{Suporte à Mobilidade em Redes Mesh Sem Fio: estratégias comuns versus SDN}

\author{Italo Brito\inst{1}, Gustavo B. Figueiredo\inst{1}\thanks{TR-PGCOMP-003/2015. Technical Report. Computer Science
Graduate Program. Federal University of Bahia.} }

\address{Departamento de Ciência da Computação -- Universidade Federal da Bahia (UFBA)\\
      Av. Adhemar de Barros, s/n -- Campus Ondina -- Salvador, BA -- Brasil
  \email{\{italo,gustavo\}@dcc.ufba.br}
}

\newcommand{\doingles}[2]{(#1, do inglês \textit{#2})}
\newcommand{\doinglesi}[2]{#1 (do inglês \textit{#2})}
\newcommand{\mesh}{\textit{mesh}\xspace}
\newcommand{\adhoc}{\textit{ad-hoc}\xspace}
\newcommand{\framework}{\textit{framework}\xspace}
\newcommand{\batman}{B.A.T.M.A.N.\xspace}
\newcommand{\openwimesh}{OpenWiMesh\xspace}
\newcommand{\backbone}{\textit{backbone}\xspace}
\newcommand{\setup}{\textit{setup}\xspace}
\newcommand{\handoff}{\textit{handoff}\xspace}

\begin{document} 

\maketitle

\begin{abstract}
Wireless mesh networks have been presented as a robust, scalable and low cost
solution to provide connectivity in long distance areas. However, given its
nature, routing strategies must support seamless mobility of nodes while
enabling operation  with a good performance and fast self-recovery from links
fault. The routing approaches that meet these requirements are based on
protocols usually used in \adhoc networks (e.g. OLSR and \batman) and, more recently,
in the SDN paradigm (e.g. OpenWiMesh), each one of them having its own pros and
cons. Therefore, it is important to evaluate the benefits and impacts of each
approach, taking into account models and metrics inherent to the mobility. In
this paper, we propose a simplified implementation of mobility support in
OpenWiMesh and evaluate its performance compared to other protocols, using
differents mobility and data traffic models. The chosen metrics are based on
packet loss, occupation of the links with control traffic and delay. Simulated
results show that the SDN approach perform better than the classic protocols,
beyond flexibility and programmability of SDN itself.
\end{abstract}

\begin{resumo}
As redes \mesh têm sido apontadas como soluções robustas, escaláveis e de baixo
custo para promover conectividade em áreas de longo alcance. Devido à natureza
dessas redes, a estratégia de roteamento deve oferecer suporte à mobilidade
transparente dos nós, além de operar com desempenho satisfatório e se auto-recuperar 
rapidamente de falhas nos enlaces que a compõe. As estratégias de roteamento que
atendem a esses requisitos baseiam-se em protocolos comumente utilizados em redes \adhoc
(e.g. OLSR e \batman) e, mais recentemente, no paradigma SDN (e.g. OpenWiMesh),
cada um com seus prós e contras. Dessa forma, é importante comparar e avaliar os
benefícios e impactos de cada estratégia levando em consideração modelos e
métricas inerentes à mobilidade. Este trabalho propõe uma implementação
simplificada de mobilidade no \framework OpenWiMesh e avalia seu desempenho
em relação à outras estratégias de roteamento, levando em consideração diferentes modelos de
mobilidade e de geração de tráfego. As métricas utilizadas incluem a taxa de
perda de pacotes, ocupação dos enlaces com tráfego de controle e atraso no
encaminhamento de mensagens. Os resultados mostram que a abordagem SDN pode
oferecer melhor desempenho que os protocolos comumente usados em redes \adhoc, além da maior
flexibilidade e programabilidade intrínsecas do paradigma.
\end{resumo}

\section{Introdução}

Uma rede em malha sem fio \doingles{WMN}{Wireless Mesh Network} é uma rede sem
fio de múltiplos saltos em que cada nó é capaz de se comunicar com todos os
demais nós da rede, seja diretamente (com seus vizinhos) ou através do
encaminhamento por nós intermediários \cite{akyildiz2009}. As WMNs têm sido
reiteradamente apontadas como a solução tecnológica de melhor custo benefício
para a construção de uma plataforma de comunicação altamente escalável e
capaz de promover conectividade a um custo relativamente baixo. Sua
utilização traz benefícios potenciais como alta capacidade de comunicação,
elevada disponibilidade e tolerância à falhas, além de propiciar uma rápida
implantação de serviços de rede.

O roteamento nas WMNs é uma área de pesquisa bastante ativa, com trabalhos
investigando diferentes
estratégias de escolha e manutenção de caminhos na rede \mesh. 
Em particular, um requisito importante dessas estratégias 
é o suporte à mobilidade dos nós, de forma que um nó móvel, ao movimentar-se, deve ter sua rota
de acesso alterada (refletindo sua nova localização) sem interrupção de
conectividade.
Algumas dessas estratégias baseiam-se em protocolos comuns de
redes \adhoc, tais como OLSR \cite{olsr2001} e \batman \cite{batman2008}, e outras,
mais recentemente, baseiam-se no paradigma de Redes Definidas por Software
\doingles{SDN}{Software Defined Networks}, como é o caso do \framework
\openwimesh \cite{openwimesh2014}.

As estratégias comuns de roteamento \adhoc estão bem consolidadas na literatura
de redes \mesh,
baseando-se essencialmente em arquiteturas distribuídas, protocolos robustos e
possibilidade de extensão, seja por modificação de parâmetros de configuração ou
do desenvolvimento de plugins. Por outro lado, pesquisas recentes mostram que
esse método de extensão é limitado \cite{dely2011}, que é necessário maior programabilidade da
rede e virtualização de recursos, permitindo assim alterações inclusive na
lógica de funcionamento das WMNs (e.g. roteamento baseado em fluxos, engenharia
de tráfego a partir de políticas administrativas, etc.). Essas questões são
então endereçadas pelo paradigma SDN, e sua aplicação em redes \mesh vem sendo
discutida recentemente na literatura. A literatura, por outro
lado, carece de estudos comparativos quanto aos benefícios e impactos de cada estratégia, 
principalmente levando em consideração modelos e métricas inerentes à mobilidade.

Nesse sentido, torna-se importante mensurar e comparar como os protocolos de
roteamento, comuns e SDN, lidam com a mobilidade na rede. Tal avaliação deve
levar em consideração a perspectiva de
prover conectividade ininterrupta aos nós móveis, sem onerar a rede com alta
taxa de ocupação por tráfego de controle ou alta latência na entrega de
mensagens. É importante, ainda, que diferentes modelagens
de padrão de movimentação e de geração de tráfego sejam consideradas, 
a fim de que os resultados sejam mais fiéis a ambientes reais.
Este artigo apresenta um estudo comparativo de estratégias de roteamento em
redes \mesh, clássicas (OLSR e \batman) e baseadas em SDN (OpenWiMesh), considerando
requisitos de mobilidade e desempenho da rede. 

O restante do trabalho está divido da
seguinte forma. A Seção \ref{sec:roteamento-wmn} apresenta uma visão geral sobre
roteamento em WMN. A Seção \ref{sec:modelagem-ambiente} discute modelos de
mobilidade e de geração de tráfego, além de apresentar o ambiente usado para
execução dos experimentos. Em seguida, na Seção \ref{sec:resultados}, os
resultados dos experimentos são apresentados e discutidos. A Seção
\ref{sec:trabalhos-relacionados} traz um breve resumo de trabalhos correlatos.
Para que, enfim, na Seção \ref{sec:conclusao} apresente-se as conclusões e
trabalhos futuros.

\section{Estratégias de Roteamento em Redes Mesh Sem Fio}
\label{sec:roteamento-wmn}

O roteamento e encaminhamento em redes \mesh sem fio é similar, de certa forma, 
ao roteamento em redes cabeadas. No entanto, alguns aspectos
inerentes ao canal sem fio como, por exemplo, a atenuação, interferência,
contenção, aliadas a
características como mobilidade e capacidade de transmissão limitada dos nós,
levam a uma alta dinamicidade da rede, que deve ser tratada pela estratégia de
roteamento adotada. Tal estratégia deve ser, portanto, otimizada a fim de
diminuir o tráfego de controle e reagir rapidamente à mudanças nos enlaces sem
fio.

As estratégias de roteamento em redes \mesh são classificadas geralmente em três
categorias: protocolos pró-ativos, reativos e híbridos. Nos protocolos
pró-ativos, computa-se as rotas para todos os nós durante a fase de
inicialização do protocolo. Já nos protocolos reativos,
as rotas são calculadas na medida em que são requisitadas pelas
aplicações. Os protocolos reativos possuem a vantagem de não ocupar
desnecessariamente a rede com
tráfego de controle, ao passo que os protocolos pró-ativos
geralmente apresentam menor atraso no estabelecimento de novos fluxos na rede,
uma vez que os caminhos já estão definidos. Os protocolos híbridos tentam
combinar as vantagens dos protocolos reativos e pró-ativos e adaptar seu
funcionamento à diferentes cenários. Mais recentemente, uma quarta categoria
passou a ser usada nesse contexto: as redes \mesh definidas por software
\cite{dely2011}. Essa é uma área de pesquisa promissora pois
combina as características de robustez e escalabilidade das redes \mesh com a
programabilidade e controle centralizado do SDN.

Um ponto que merece especial atenção na escolha da estratégia de roteamento é
o suporte à mobilidade transparente entre os nós, onde aplicações de tempo real
não tem sua conectividade afetada em face a movimentação de um nó. Nesse
sentido, dois requisitos são de importância destacável: mudança de ponto de acesso (processo
também conhecido como \handoff) sem interrupção de conectividade
\cite{amir2006fast}, e propagação da atualização de rotas depois do \handoff. Este
trabalho aborda principalmente os aspectos de atualização de rotas após o
\handoff.

Nesta seção serão apresentadas duas estratégias comumente usadas para roteamento em WMN
e também uma estratégia baseada em SDN, as quais serão alvo de
avaliação de desempenho deste trabalho. As estratégias clássicas usadas foram o
OLSR e \batman, pois são as mais referenciadas pela literatura
\cite{mishra2013,abolhasan2009real,akyildiz2009}. Já para a estratégia SDN,
adotamos o OpenWiMesh \cite{openwimesh2014}, que já possui algoritmos de 
engenharia de tráfego na rede \mesh e foi estendido para suportar, de forma
simples, mobilidade dos nós.

\subsection{OLSR}

O protocolo \doinglesi{OLSR}{Optimized Link State Routing} \cite{olsr2001} é um
protocolo de roteamento pró-ativo, baseado em algoritmo otimizado de estado do
enlace e amplamente utilizado em redes \mesh.
A principal vantagem do OLSR em relação aos algoritmos de estado de enlace
clássicos é a existência de ``retransmissores multiponto''
\doingles{MPRs}{multipoint relay}. Os MPRs são nós especialmente selecionados para
fazer a transmissão das informações de topologia, evitando que o processo
de inundação seja realizado a partir de todos os nós. Esta técnica minimiza a contenção e o
número de mensagens de controle necessárias para estabelecimento da tabela de
roteamento.

O protocolo faz uso de mensagens HELLO periódicas
para descoberta de vizinhos e sinalização da seleção de MPRs. Além disso,
utiliza também mensagens 
TC (controle de topologia) para troca de informações de topologia entre nós.
Todos os nós, assim, conhecem a topologia e calculam localmente as melhores
rotas para os demais nós.

O OLSR trata a mobilidade a partir do recálculo e atualização das rotas de saída e
entrada de um nó.
Para as rotas de saída, deve-se definir qual
vizinho será o novo gateway para cada um dos possíveis destinos na rede, enquanto que para as rotas de entrada é necessário
atualizar a tabela de roteamento dos demais nós da rede com o novo caminho
através do qual o nó móvel é alcançável. A atualização das rotas
de saída é baseada nas mensagens de HELLO, que são enviadas a cada
$ t_{hello} $ segundos. Para viabilizar mobilidade transparente, é preciso que o
valor de $t_{hello}$ seja baixo (e.g.
500ms \cite{amir2006fast}), porém mesmo nessa configuração podem ser necessários múltiplos ciclos de mensagens
HELLO para estabelecer enlaces bidirecionais.
A atualização das rotas de entrada depende da propagação das informações de
topologia, que ocorre através das mensagens de TC a cada
$t_{tc}$ segundos, e do recálculo da tabela de rotas em todos os nós.

Em ambos os casos, $t_{hello}$ e $t_{tc}$
poderiam ser reduzidos a grandezas muito baixas para permitir mobilidade
transparente, porém dessa forma o tráfego de controle traria alto impacto à
rede devido a contenção para controle de colisão.

\subsection{\batman}

O protocolo \doinglesi{\batman}{Better Approach to Mobile Ad-Hoc Networking} 
\cite{batman2008} também faz uso de uma estratégia pró-ativa,
todavia a técnica de distribuição da topologia e seleção de rotas é melhor
alinhada com a limitação de recursos de hardware geralmente disponíveis nos nós da rede
\mesh sem fio.
No \batman, os nós não conhecem toda a topologia da rede. Ao invés disso, eles
conhecem apenas o vizinho que possui a melhor rota para determinado destino.
Adicionalmente, o algoritmo de seleção de rotas baseia-se no fato de
que os melhores enlaces entregarão as mensagens de forma mais rápida e
confiável. Assim, a seleção de melhores rotas se dá de forma natural, sem
necessidade de um cálculo baseado na topologia em cada nó.

De forma geral, cada nó envia, por difusão, \textit{mensagens de originadores}
\doingles{OGMs}{originators messages}, informando aos vizinhos sobre a sua
existência. Ao receber uma mensagem OGM, o nó retransmite essa mensagem para
seus vizinhos. 
Naturalmente, espera-se que as mensagens 
encaminhadas por enlaces de maior qualidade cheguem primeiro, por isso 
a primeira mensagem recebida torna-se o melhor caminho para o originador daquela
mensagem e as demais são descartadas. 
Apenas a primeira OGM recebida para determinado destino (melhor caminho) é
retransmitida.
Existe, portanto, uma inundação da rede com OGMs.
Ademais, uma checagem bidirecional de cada
enlace é feita a fim de verificar se o enlace detectado pode ser usado em ambas
as direções.

As mensagens OGM são usadas pelo \batman tanto para detectar mudanças quanto
qualidade dos enlaces. Quando um nó se move na rede, seus vizinhos
levam no mínimo dois ciclos de OGM para atualizar as tabelas de rota, enquanto
que o restante da rede depende da difusão dessas OGMs para propagar essa
atualização \cite{batman2008}.
Isso implica em tempos razoáveis para atualização das rotas de saída, mas em
desempenho ruim para rotas de entrada.

Uma implementação bastante utilizada do \batman é o BATMAN-Adv (\batman
\textit{Advanced}) \cite{batmanadv}, que opera em camada 2, possui mecanismos de
controle de loop e diversas outras funcionalidades.

\subsection{\openwimesh}

O \openwimesh é um \framework para roteamento e engenharia de tráfego em redes
\mesh sem fio, através de uma abordagem baseada em Redes Definidas por Software
(SDN) \cite{openwimesh2014}. O \openwimesh implementa o roteamento na rede \mesh a partir de uma abordagem
centralizada, alinhada com o conceito de SDN, definindo caminhos sob demanda a
partir de um grafo que modela a rede. Valendo-se da programabilidade da rede
proporcionada pelo SDN, o OpenWiMesh permite a definição de diversas estratégias para
escolha dos melhores caminhos. 
Em particular, três estratégias são
pré-distribuídas em conjunto com o \openwimesh: \doinglesi{HC}{Hop-Count}, que defini a melhor
rota baseado na menor quantidade de saltos; \doinglesi{HLRB}{Highest Link
Residual Bandwidth}, onde o melhor caminho é escolhido como aquele cuja banda
residual é maior; e \doinglesi{HLRB-SHC}{Highest Link Residual Bandwidth in Same
Hop Count}, que combina as duas estratégias anteriores a fim de escolher caminhos
menos congestionados porém sem aumentar significativamente a quantidade de saltos.
Neste trabalho foi utilizado algoritmo baseado em contador de saltos.

O grafo da rede é mantido de forma centralizada em um nó com maior capacidade
computacional definido como 
controlador, responsável pelo cálculo de 
rotas e gerencia da rede \mesh. Dessa maneira os demais nós da rede \mesh não 
são onerados com funções de cálculo de rotas ou armazenamento da topologia.
Pelo contrário, 
cada nó de \backbone possui apenas um componente local, chamado de
\textit{graphClient}, cuja função é coletar informações do
meio sem fio (lista de vizinhos, potência de sinal de cada vizinho,
SINR, etc) e enviá-las ao controlador. O controlador então atualiza o grafo da
rede, que será usado para definição das rotas futuras. 

O funcionamento do \openwimesh pode ser divido em duas fases: uma fase de \setup
da rede e uma fase de criação e manutenção de rotas para aplicações.
Na fase de \setup os nós tem sua conexão estabelecida com o controlador, através
do protocolo Openflow \cite{mckeown2008} em modo \textit{in-band}.
Já na fase de criação e
manutenção de rotas de aplicações, os nós \mesh enviam, sobre demanda,
requisições de encaminhamento ao controlador (\textit{Packet-In}), que faz uma
busca no grafo da rede a partir da estratégia de roteamento definida.

\begin{figure}[h]
\begin{center}
\includegraphics[width=.8\textwidth]{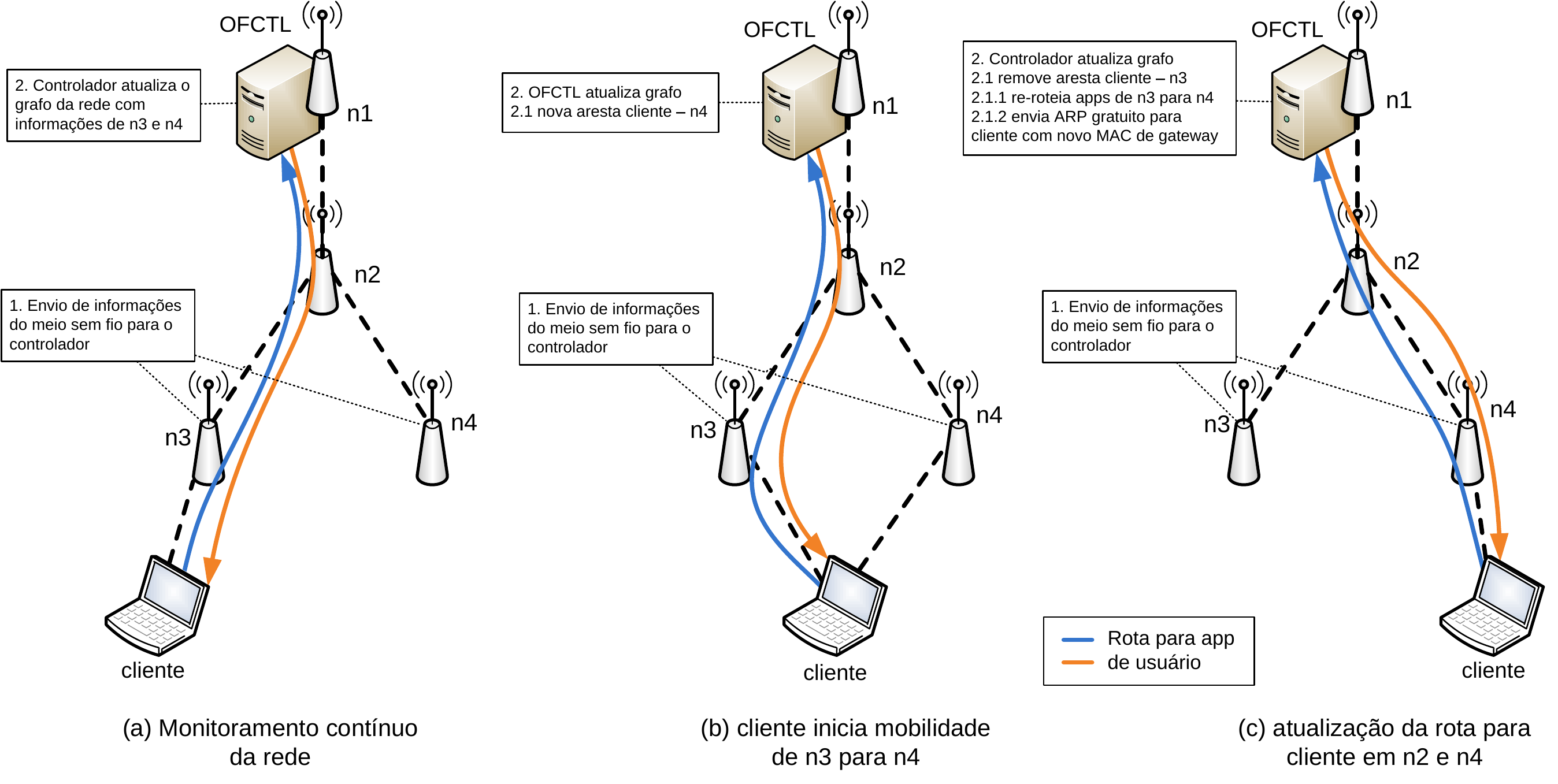}
\caption{Gerenciamento de mobilidade no \openwimesh}
\label{fig:gerenciamento-mobilidade-openwimesh}
\end{center}
\end{figure}

Na versão original do \openwimesh não havia tratamento para nós móveis, sendo
necessário que o próprio nó desse início à reconexão após o \textit{timeout} do
sistema, gerando por vezes mais de 30 segundos de desconexão. 
Neste trabalho foi desenvolvida uma extensão do \framework para adicionar
gerenciamento de 
mobilidade dos nós. A extensão
proposta, conforme Figura \ref{fig:gerenciamento-mobilidade-openwimesh}, 
atua na fase de criação e manutenção de rotas de aplicações,
onde o controlador monitora a lista de informações do meio sem fio enviada pelo
\textit{graphClient}, e verifica se houveram desassociações cujo enlace
estava em uso por alguma aplicação. Nesse caso é iniciada uma ação de
re-roteamento na rede, atualizando a tabela Openflow dos nós que utilizavam
aquela aplicação (rota de
entrada) e enviando uma resposta ARP gratuita ao nó móvel (rota de saída). 
O envio de mensagem ARP gratuita é importante para que o nó móvel atualize sua
tabela ARP substituindo a(s) ocorrência(s) do MAC do nó com o qual ele estava associado
anteriormente para o MAC do novo nó de associação. Dessa maneira o
\textit{handoff} do nó móvel é gerenciado pelo \framework OpenWiMesh e não pelo nó
que está se movendo, viabilizando a implementação de diversas estratégias de
detecção e gerenciando de \textit{handoff} pelo controlador da rede.

Os nós móveis podem ser clientes e até mesmo nós de \backbone.
Caso o nó móvel seja um nó de \backbone,
apenas atualizar o MAC do controlador é suficiente, pois as demais entradas da
tabela ARP direcionam para o MAC daquele próprio nó (as regras Openflow
reescrevem o MAC para o próximo salto). Porém em caso de tratar-se de nó cliente, serão enviadas
tantas respostas ARP gratuitas quanto forem necessárias para aplicações que utilizavam
aquele enlace. A gerencia de mobilidade, dessa maneira, propaga as atualizações de
rota de forma rápida e eficiente.

O código-fonte do OpenWiMesh, bem como a extensão desenvolvida neste trabalho,
estão disponíveis no site do projeto \cite{owmsite}.

\section{Trabalhos relacionados}
\label{sec:trabalhos-relacionados}

Alguns trabalhos estão disponíveis na literatura avaliando protocolos de
roteamento em redes \mesh, porém nenhum deles compara abordagens clássicas com
abordagens SDN. O trabalho \cite{mishra2013} compara os efeitos de diferentes
modelos de mobilidade e geração de tráfego com base em protocolos de redes \adhoc
AODV, OLSR e ZRP, levando em consideração métricas como atraso fim-a-fim, jitter
e vazão. Embora faça uso de dois modelos de mobilidade, os impactos da
mobilidade na conectividade dos nós não são avaliados. Naquele trabalho faz-se
uso de simulação, através da ferramenta QualNet.

Já o trabalho \cite{abolhasan2009real} investiga o desempenho de protocolos
de roteamento pró-ativos em ambiente real, com testbed de rede \adhoc construído
em área fechada. Os protocolos comparados são OLSR, \batman e BABEL. As métricas
daquela avaliação foram vazão, taxa de entrega de pacotes e latência para
recuperação de falhas em enlaces. Este trabalho não leva em consideração a
característica de mobilidade das redes \adhoc. Os resultados daquele trabalho 
estão alinhados com aqueles aqui apresentados:
o \batman possui menor tempo de recuperação de falhas em enlaces que o OLSR (que aqui é
representado pela menor taxa de perda de pacotes).

O trabalho \cite{dely2011} foi o primeiro a integrar Openflow em WMNs,
propondo uma arquitetura que viabilize tal integração e provendo a
funcionalidade de roteamento baseado em fluxos. Como estudo de caso foi
implementado uma solução de mobilidade simplificada entre nós \mesh e avaliado
em uma rede real de testbed (KAUMesh). Uma das métricas avaliadas foi a
sobrecarga do tráfego de controle, comparando OLSR e a proposta Openflow, a
medida que novos fluxos são iniciados. O resultado daquela métrica mostrou que o
Openflow apresenta crescimento linear com a quantidade de novos fluxos por
segundo, enquanto o OLSR permanece constante (característica esperada dado sua
configuração de rotas pró-ativa). Outra métrica avaliada naquele trabalho foi o
tempo de desconexão (\textit{handover}) do nó \mesh móvel. Alcançou-se, naquela
ocasião, um tempo de interrupção de 210ms na média, com variação entre 50ms e
270ms. Não estão disponíveis informações sobre o modelo de
mobilidade utilizado, enquanto que o padrão de tráfego considera um intervalo
entre pacotes de 10ms (100pps) e transporte UDP, via ferramenta \textit{iperf}.

\section{Cenário de Avaliação}
\label{sec:modelagem-ambiente}


\subsection{Modelos de Mobilidade}

\begin{figure}[htbp]
\centering
\setlength{\unitlength}{0.0105in}
\mbox{
\subfigure[Modelo \textit{Random Waypoint}]{\includegraphics[width=0.5\textwidth]{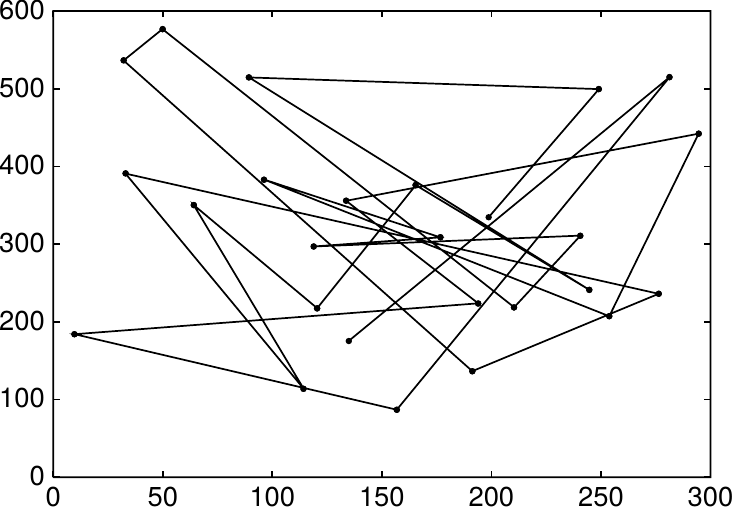}}
\subfigure[Modelo \textit{Reference Point Group Mobility}]{\includegraphics[width=0.5\textwidth]{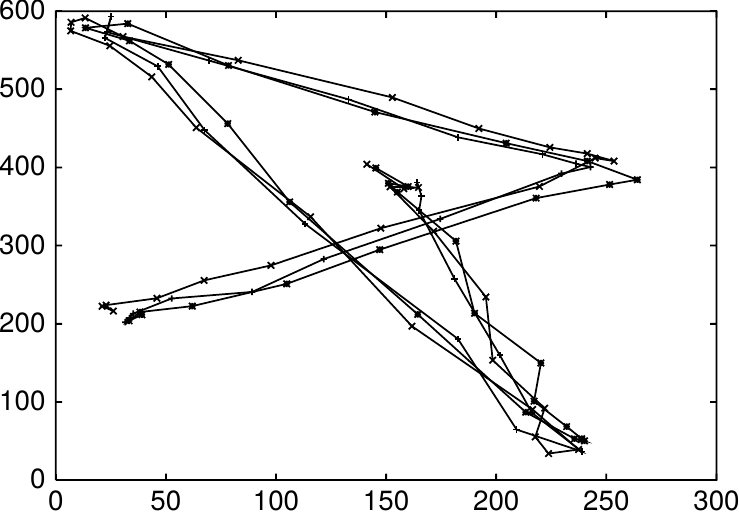}}
}
\caption{Padrão de movimentação por modelo de mobilidade \cite{camp2002survey}}
\label{fig:mobmodel-rwp-rpgm}
\end{figure}

O modelo de mobilidade define o padrão de movimentação dos nós na rede.
Existem basicamente dois tipos de modelos de mobilidade: mobilidade de
entidade ou individual e mobilidade em grupo \cite{camp2002survey}. 
Para o tipo de mobilidade
individual existem diversos modelos disponíveis, tais como \textit{Random
Waypoint}, \textit{Random walk}, \textit{Gauss-Markov}, etc. Já na mobilidade
em grupo pode-se citar os modelos \textit{Exponential Correlated Random
Mobility}, \textit{Nomadic Community Mobility} e, mais utilizado,
\textit{Reference Point Group Mobility Model}. Neste trabalho serão
utilizados os modelos \textit{Random Waypoint (RWP)} e \textit{Reference Point
Group Mobility Model (RPGM)}, ilustrados na Figura \ref{fig:mobmodel-rwp-rpgm} 
e detalhados a seguir.

O modelo RWP é um modelo de mobilidade simples, onde o padrão de movimentação de
um nó é modelado como a escolha de um destino aleatório dentro da área
de execução e de uma velocidade uniformemente distribuída entre
\verb=[minspeed, maxspeed]=. Cada nó pode ainda pausar em determinado
local por certo período de tempo. O comportamento do modelo RWP pode ser visto
na Figura \ref{fig:mobmodel-rwp-rpgm}{a}.
Este é um modelo de mobilidade amplamente 
utilizado em estudos de redes \adhoc \cite{abolhasan2009real,mishra2013}. Por
vezes esse modelo é simplificado, eliminando-se, por exemplo, os períodos de
pausa, abordagem adotada neste trabalho. Alguns aspectos importantes no uso
deste modelo são seus parâmetros de configuração (velocidades mínima e máxima, período de
pausas, área), a disposição inicial dos nós na área de execução e período de
execução do experimento \cite{camp2002survey}. 
Acerca da disposição inicial dos nós, neste trabalho os
nós foram distribuídos uniformemente na área de execução. Em se tratando do
período de execução, recomenda-se ignorar os primeiros segundos devido a alta
variação no grau de conectividade dos nós. Neste trabalho foram ignorados os
primeiros 900 segundos do modelo de mobilidade.

Já o RPGM representa a movimentação de um grupo de nós em determinada direção,
bem como uma movimentação individual dentro do grupo. O RPGM é apropriado
para modelar situações de desastres, por exemplo, onde grupos de pessoas se movimentam
juntas em determinada direção (saídas de emergência) ou ainda em cenários de
cooperação mútua \cite{camp2002survey}. Os movimentos do grupo são guiados por
um ponto lógico central do
grupo, chamado de vetor de movimento \verb=GM=. Este vetor de movimento
caracteriza o movimento de todo o grupo, incluindo seu destino e velocidade.
Dentro do grupo, os nós se movem aleatoriamente em relação a seu ponto
referencial. A Figura \ref{fig:mobmodel-rwp-rpgm}{b} ilustra a movimentação de
três nós com o modelo RPGM. Novamente, a posição inicial dos nós, composição dos
grupos e velocidade, são configurações importantes para o modelo. Neste trabalho adotou-se
uma distribuição uniforme dos nós na área de execução (assim como no modelo RWP)
e grupos de tamanho médio de três nós.

\subsection{Modelos de Tráfego}

Para avaliar as estratégias de maneira mais realística e sobre condições 
extremas, foram utilizados modelos estocásticos para representar 
o padrão de geração de tráfego na rede
durante o período de execução dos testes. Nesse trabalho fez-se uso de modelos de
geração de tráfego baseados em taxa de bits constante \doingles{CBR}{Constant
Bit Rate} e taxa de bits variável \doingles{VBR}{Variable Bit Rate}. Em
particular, duas variáveis aleatórias foram utilizadas para caracterizar o modelo de
tráfego: o tamanho do pacote \doingles{PS}{Packet Size} e o intervalo entre o envio
de pacotes \doingles{IDT}{Inter Departure Time}.

%

\subsection{Ambiente de Avaliação}

Neste trabalho optou-se pela experimentação
através de emulação devido a dificuldade de montagem de ambiente real,
isolamento de efeitos de interferência externos e 
replicabilidade. A ferramenta de emulação utilizada foi o 
CORE \cite{ahrenholz2010comparison}, que faz uso de recursos de virtualização do
kernel do Linux (\textit{network namespaces}) em conjunto com um modelo de rede
sem fio 802.11 simplificado.

Os experimentos foram conduzidos tomando-se como fatores a topologia da rede
\mesh, modelos de geração de tráfego e modelos de mobilidade. A respeito das
topologias, foram criados cenários com 10, 20 e 30 nós, onde o diâmetro da rede
em cada um deles foi 2, 4 e 6 saltos, respectivamente. 
Houve também uma preocupação de manter
o mesmo grau de conectividade dos nós, definindo, para isso, áreas de
execução com tamanho proporcional à quantidade de nós. Em
todas as topologias foram usados 50\% de nós de \backbone estacionários e 50\% de nós
clientes móveis. Os nós de \backbone foram posicionados para que houvesse cobertura
total da área de execução. O cálculo do diâmetro e grau considera 
apenas os nós de \backbone.

Os modelos de mobilidade adotados foram RWP e RPGM. O cenário de mobilidade teve
duração de 120 segundos, iniciados após a convergência do protocolo de
roteamento e um período monótono (sem tráfego e sem mobilidade) de 60s. 
Os nós foram configurados para movimentar-se a uma velocidade entre
2m/s e 6m/s, sem período de pausa (cenário com ampla mobilidade). A ferramenta
utilizada para gerar os scripts de mobilidade foi o \textit{BonnMotion}
\cite{bonnmotion2010}.

No que tange ao modelo de geração de tráfego, para o CBR
tanto o PS quanto o IDT foram mantidos constantes, fixados em 50 pps e 1400
bytes respectivamente, ao passo que no VBR
adotou-se uma distribuição uniforme para o PS, variando entre 64 e 1400 bytes, e
uma distribuições de Poisson para IDT com média de 50 pps. Em ambos os modelos
foi gerado um fluxo de dados UDP a partir de um nó central 
da topologia (simulando um nó de \backbone com acesso à Internet) para
todos os outros nós móveis. A ferramenta utilizada como gerador de tráfego
foi o D-ITG (\textit{Distributed Internet Traffic Generator}) \cite{ditg2004}.
A Tabela \ref{table:emulation-parameters} apresenta um resumo da configuração do
ambiente de execução.

\begin{table}
\begin{center}
\label{table:emulation-parameters}
\caption{Parâmetros de emulação}
\begin{tabular}{| p{6cm} | p{8cm} |}
\hline
\textbf{Parâmetro de emulação} & \textbf{Valor} \\ \hline
\multirow{3}{*}{Topologias} & T1: 10 nós, diâmetro 4, dimensão 240x360m \\ & T2: 20 nós, diâmetro 6, dimensão 400x360m \\ & T3: 30 nós, diâmetro 8, dimensão 560x360m \\ \hline
Disposição dos nós na topologia & Uniformemente distribuídos \\ \hline
Modelo de mobilidade & RWP e RPGM (ambos sem pausa) \\ \hline
Modelo de Tráfego & CBR e VBR \\ \hline
Tempo de execução & 180s (+ tempo de convergência dos protocolos) \\ \hline
Modelo de perda de propagação & \textit{Free Space Path Loss} \\ \hline
Modelo de antena & Omni-directional\\ \hline
Sensibilidade de recepção & -90dBm\\ \hline
Potência de transmissão & 20dBm\\ \hline
Ganho da Antena & TX 1 dBi e RX 1 dBi\\ \hline
Camada física & IEEE 802.11ag (canal 11, frequência 2.4 GHz)\\ \hline
\end{tabular}
\end{center}
\end{table}

As topologias, modelos de mobilidade, scripts dos experimentos e outros códigos
estão disponíveis no repositório do projeto OpenWiMesh
(\url{https://nuvem.pop-ba.rnp.br/gitlab/ufba/openwimesh}).

\section{Análise dos Resultados}
\label{sec:resultados}

O estudo comparativo entre as estratégias de roteamento em WMN deu-se
pela avaliação do OLSR, BATMAN-Adv e \openwimesh com relação às métricas i)
\textbf{perda de pacotes}, tomando como fatores a topologia (quantidade de nós,
diâmetro da rede), modelo de tráfego e modelo de mobilidade; ii) \textbf{tempo
de resposta} (latência), baseado no tempo de ida e volta (RTT) tanto para o pior
caso (\textit{slowpath}) quanto para o caso médio (\textit{fastpath}); e iii) \textbf{ocupação dos
enlaces com tráfego de controle}.


Todos os experimentos foram repetidos 30 vezes. Cada execução durou cerca de 240
segundos. As três estratégias aqui avaliadas foram configuradas de forma
equivalente, com intervalo entre atualizações de 1 segundo.  O intervalo de
confiança foi calculado considerando-se um 
nível de confiança de 95\%.


\begin{figure}[h]
\begin{center}
\includegraphics[width=0.6\textwidth]{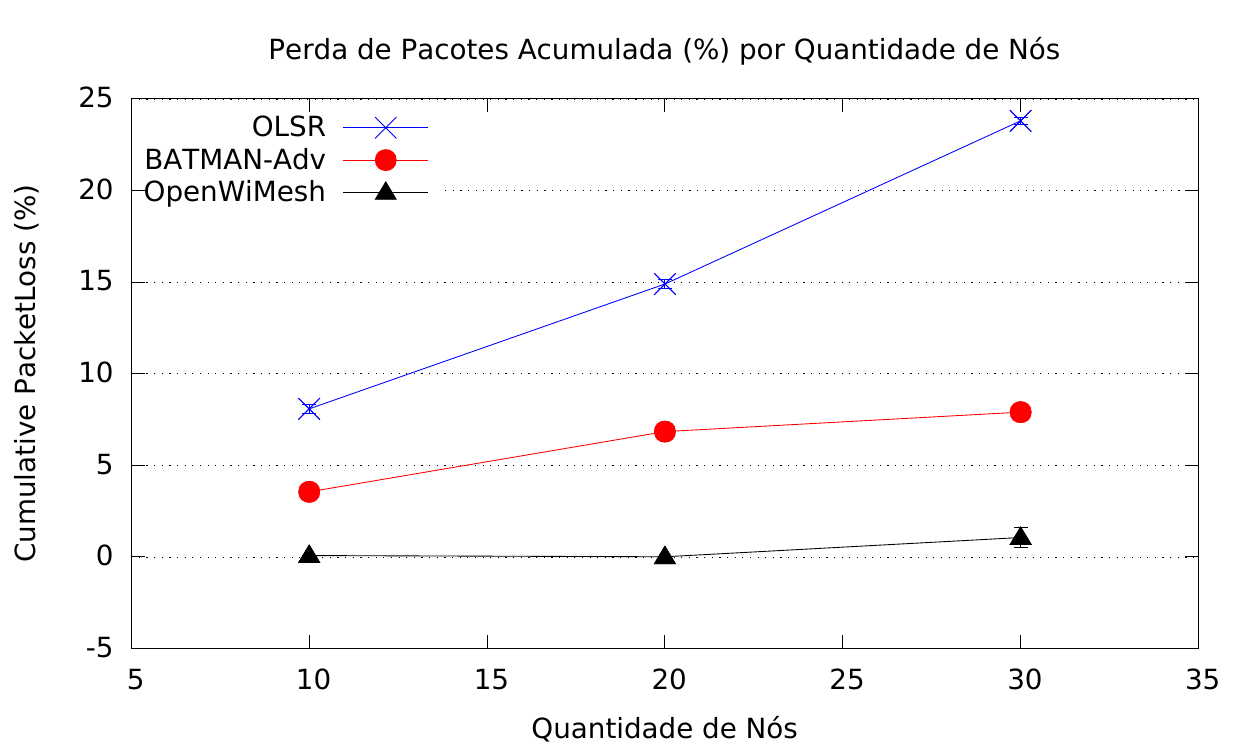}
\caption{Perda de Pacotes variando-se a Topologia}
\label{fig:pktloss-qtdnodes}
\end{center}
\end{figure}

A primeira métrica avaliada foi a perda de pacotes durante a movimentação dos
nós pela rede. Vale reforçar que todas as topologias foram configuradas
para que não houvesse áreas sem cobertura. A Figura \ref{fig:pktloss-qtdnodes}
apresenta o comportamento dos protocolos com a variação das topologias,
mantendo fixos o modelo de mobilidade RWP e de tráfego CBR. Os protocolos
BATMAN-Adv e OLSR apresentaram elevado grau de perda de pacotes, chegando a
aproximadamente 8 e 25\% de perda, respectivamente, para topologias com 30 nós.
Já o OpenWiMesh manteve um nível de perda de pacotes baixo, próximo a 0\% nas topologias
com 10 e 20 nós e atingindo um intervalo de confiança de \verb=[0.51, 1.59]= na 
topologia com 30 nós, com nível de significância de 0.05. Estes resultados
refletem a dificuldade de propagação da atualização de rotas para os demais nós da rede,
principalmente no caso do OLSR que depende da sincronização das informações de
topologia por todos os nós. O \batman, por não necessitar que todo nó
conheça a topologia completa, tem desempenho melhor que o OLSR, porém ainda
assim depende da propagação das OGMs para atualizar as rotas de entrada. Por
outro lado, o \openwimesh consegue responder de forma mais rápida à mobilidade,
devido a sua gerência centralizada e informações do grafo da rede.

\begin{figure}[h]
\begin{center}
\includegraphics[width=0.6\textwidth]{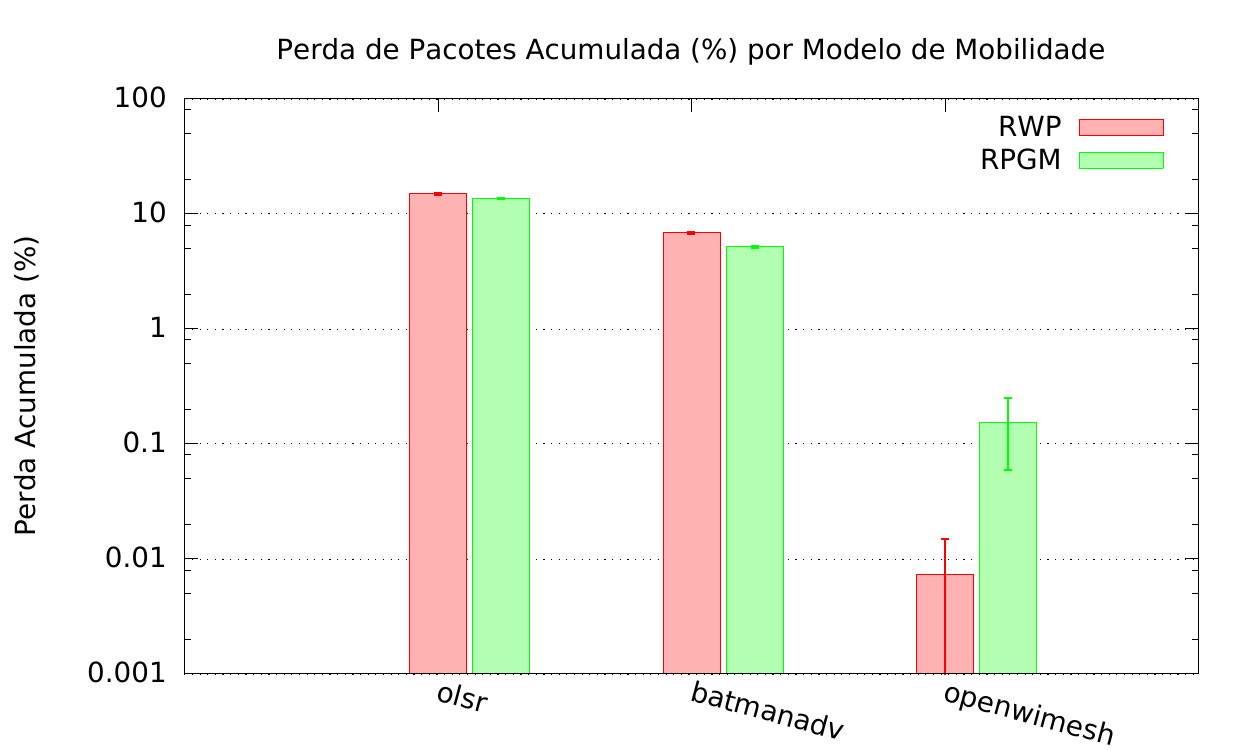}
\caption{Perda de Pacotes variando-se o Modelo de Mobilidade}
\label{fig:pktloss-mobmodel}
\end{center}
\end{figure}

Avaliou-se também o impacto dos modelos de mobilidade na entrega de pacotes,
conforme Figura \ref{fig:pktloss-mobmodel}. Pode-se observar que o OLSR
apresenta significativa perda de pacotes tanto com RWP quanto com RPGM,
divergindo em apenas cerca de 1\%. O BATBAM-Adv também apresenta perda de
pacotes significativa e também com baixa divergência entre os modelos de
mobilidade. Chama atenção, todavia, a diferença visual dos resultados do
OpenWiMesh. No entanto, observando os valores do intervalo de confiança, essa diferença
é de apenas 0.04\%, com ambos os modelos gerando uma perda máxima de 0.3\%. Para
essa métrica foi utilizada a topologia com 20 nós e modelo de tráfego CBR.

\begin{figure}[h]
\begin{center}
\includegraphics[width=0.6\textwidth]{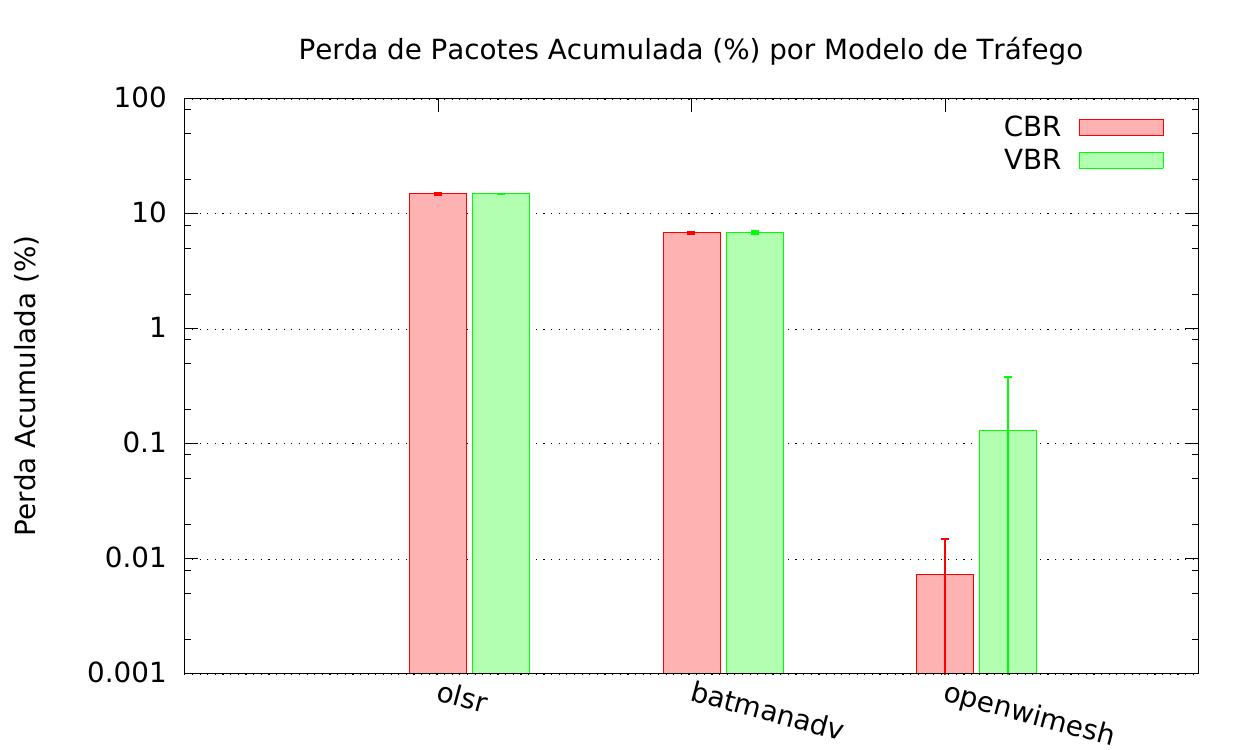}
\caption{Perda de Pacotes variando-se o Modelo de Tráfego}
\label{fig:pktloss-trafmodel}
\end{center}
\end{figure}

A variação do modelo de tráfego também não impôs significativa diferença
em relação ao cenário anterior, conforme pode ser visto na Figura
\ref{fig:pktloss-trafmodel}.


Outra métrica avaliada foi o tempo de resposta na transmissão de dados,
sendo utilizada a topologia com 20 nós e empregando o modelo de mobilidade RWP. O tempo
de resposta levou em consideração o tempo de ida e volta do pacote,
analisando o tempo do primeiro pacote, também conhecido como \textit{slowpath}, e
a média dos demais pacotes, chamado de \textit{fastpath}. O primeiro pacote é
considerado \textit{slowpath} pois ele pode necessitar de alguma configuração do
protocolo de roteamento, antes que o fluxo esteja apto a ser encaminhado. Esse é
um comportamento normal de protocolos sobre demanda, como o OpenWiMesh, porém 
mesmo o BATMAN-Adv pode ter um \textit{slowpath} significativo devido a
resolução ARP no sistema operacional. 

\begin{figure}[h]
\begin{center}
\includegraphics[width=0.6\textwidth]{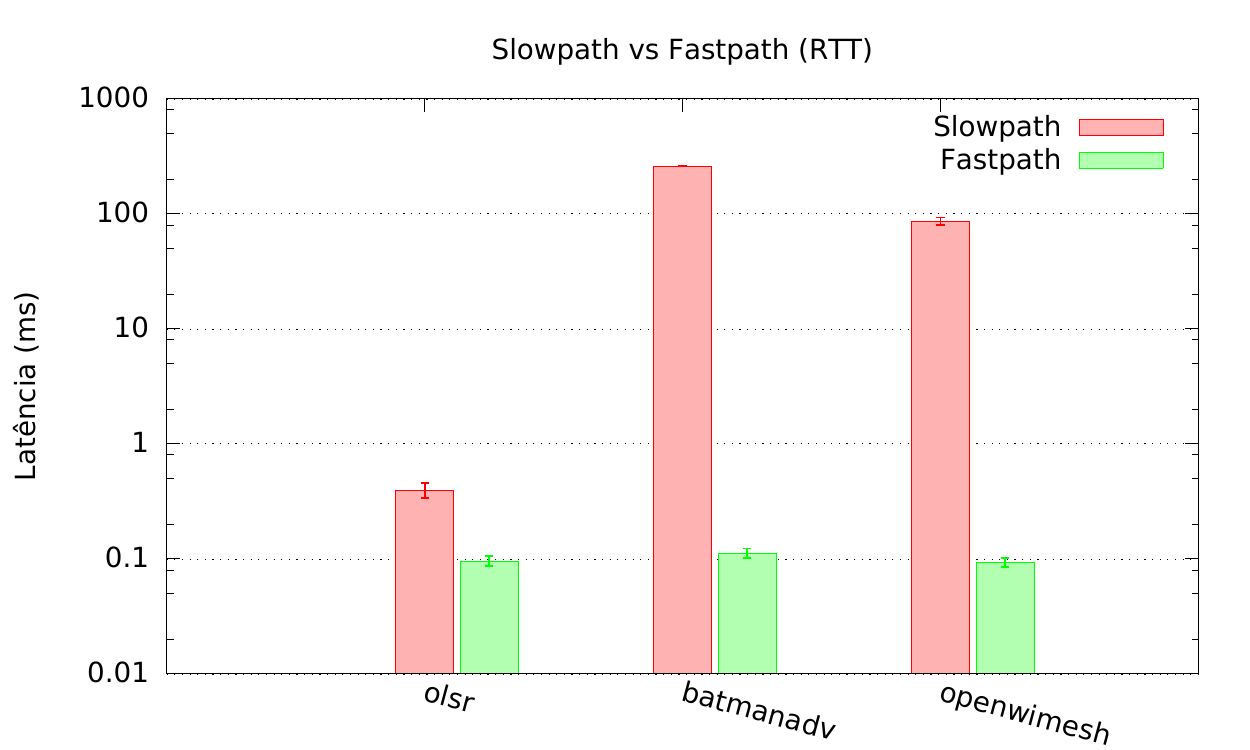}
\caption{Tempo de Resposta - \textit{Slowpath} vs \textit{Fastpath}}
\label{fig:slowfastpath}
\end{center}
\end{figure}

A Figura \ref{fig:slowfastpath} mostra o resultado da avaliação do tempo 
de resposta. O resultado do \textit{fastpath} apresenta um equilíbrio entre
as estratégias, com tempos de resposta de aproximadamente 0.01 ms, com intervalo
de confiança de 0.01 ms e nível de confiança de 95\%. Esse resultado, por outro
lado, precisa ser investigado levando-se em consideração outros fatores, como
caminhos congestionados e considerando múltiplos fluxos simultâneos na rede. Já
o resultado do \textit{slowpath} apresenta uma vantagem do OLSR que consegue
encaminhar o primeiro pacote de um fluxo com atraso menor que 1 ms. O resultado
do BATMAN-Adv, embora sobremaneira elevado, com 259 ms (intervalo de confiança
0.9ms), representa o comportamento
convencional da maioria dos protocolos de roteamento, que depende de uma
resolução ARP. Nesse sentido, o OpenWiMesh apresenta um resultado
significativamente melhor, com 86ms (intervalo de confiança de 6ms). Além do
tempo menor, um resultado indireto do OpenWiMesh sobre os demais é que a
requisição ARP não é enviada em difusão para os outros nós da rede. Pelo 
contrário, ela é captura logo na origem e respondida pelo controlador.


\begin{figure}[h]
\begin{center}
\includegraphics[width=0.6\textwidth]{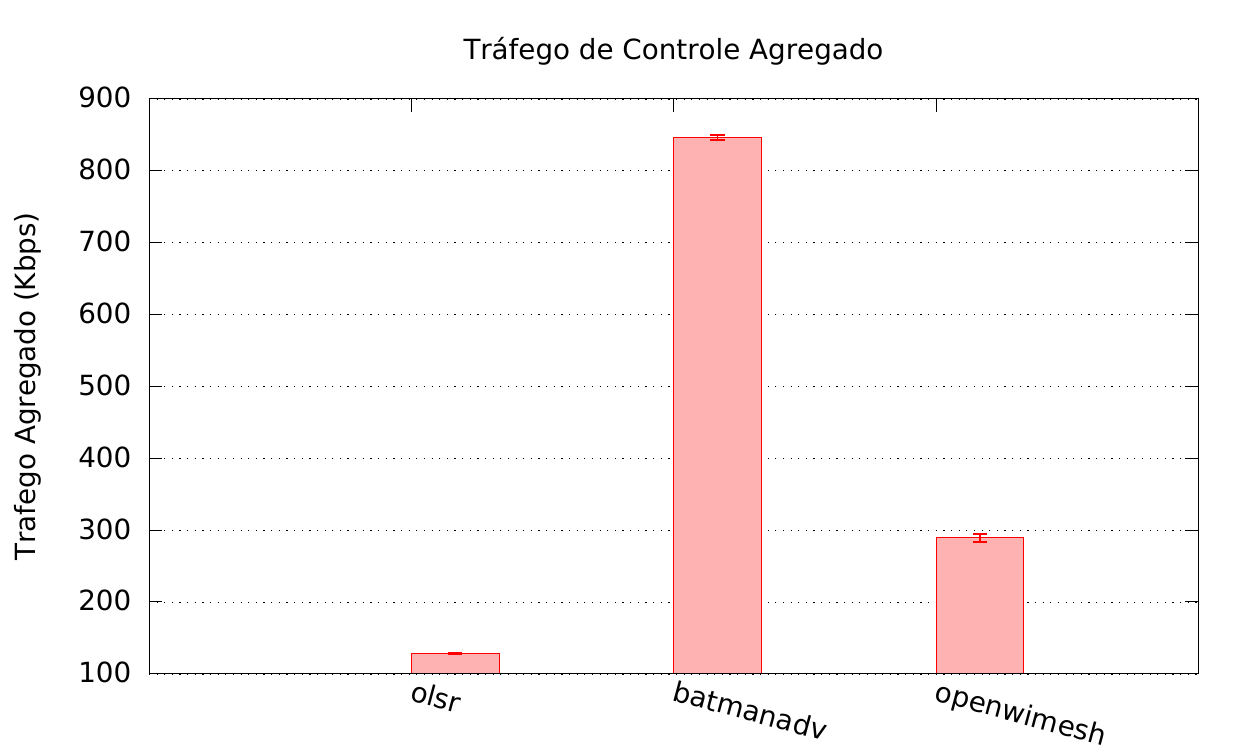}
\caption{Tráfego de Controle Agregado}
\label{fig:ctltraff}
\end{center}
\end{figure}

Neste experimento o objetivo é avaliar qual a medida de tráfego de controle que
é gerado pelos protocolos de roteamento. Idealmente a rede não deve ser
significativamente ocupada pelo tráfego de controle. O protocolo de roteamento
deve encontrar um equilíbrio entre detectar rapidamente movimentos ou falhas nos
enlaces e nós, porém sem sobrecarregar a rede. Os resultados dessa métrica estão dispostos
na Figura \ref{fig:ctltraff}, que apresenta o tráfego agregado de todos os nós
da topologia (20 nós) antes que o gerador de tráfego injete pacotes na rede.
O OLSR apresenta baixa taxa de ocupação da rede com controle e o OpenWiMesh
apresenta uma média de tráfego bem aceitável, com valor agregado de
aproximadamente 290 Kbps (intervalo de confiança de 5 Kbps). O BATMAN-Adv, por
outro lado, apresentou uma significativa taxa de ocupação da rede com tráfego
de controle. Em um ambiente real isso pode gerar contenção, podendo
impactar negativamente nas aplicações da rede.

Este trabalho não considera requisições de rotas (ativação de novas aplicações)
na computação da sobrecarga do tráfego de controle.

\section{Conclusões e Trabalhos Futuros}
\label{sec:conclusao}

Este trabalho apresentou um estudo comparativo de estratégias de roteamento em
redes \mesh, considerando abordagens comumente usadas como OLSR e BATMAN e uma
abordagem SDN (OpenWiMesh). O comparativo levou em consideração requisitos como mobilidade e
desempenho da rede. Os resultados mostram que o OpenWiMesh apresenta menor
perda de pacotes com a mobilidade de nós, com taxas muito próximas a 0\% de
perda. Tal sucesso na entrega de pacotes não implica em maior oneração para a
rede, uma vez que o tráfego de controle apresentou limites aceitáveis, comparado
com os outros protocolos e com a capacidade de transmissão dos nós, 
e a latência no encaminhamento de pacotes esteve
estatisticamente igual aos demais protocolos considerando o \textit{fastpath}.
Já a latência do \textit{slowpath}, apesar de alta, está abaixo do tempo de
estratégias baseadas em requisições ARP, como o BATMAN-Adv. Além disso, a
resolução ARP do OpenWiMesh é mais eficiente que as abordagens convencionais, uma
vez que evita que a requisição seja difundida por toda a rede
(\textit{broadcast}).

Como proposta futuras, este trabalho pode ser estendido para contemplar também a avaliação em ambientes
reais, considerando aspectos como contenção e interferência do canal sem fio,
além de incluir novas métricas como a sobrecarga do tráfego de controle face a
requisições simultâneas de novos fluxos e a distribuição de carga na rede.
Convém, ainda, investir na otimização dos protocolos utilizados para cenários de
alta mobilidade e experimentar técnicas de gerenciamento de \handoff.

\bibliographystyle{sbc}
\bibliography{comp-wmn-proto.bib}

\end{document}